\documentclass{aa}
\usepackage{epsfig}

\def\bi{\bibitem{}}

\def\beb{\begin{thebibliography}{999}}
\def\eeb{\end{thebibliography}}
\def\bei{\begin{itemize}}
\def\eei{\end{itemize}}
\def\bef{\begin{figure}}
\def\eef{\end{figure}}
\def\ben{\begin{enumerate}}
\def\een{\end{enumerate}}
\def\beq{\begin{equation}}
\def\eeq{\end{equation}}
\def\ber{\begin{eqnarray}}
\def\eer{\end{eqnarray}}

\begin{document}

\title{Scatter broadening of pulsars in the direction of the Gum nebula}
\author{D. Mitra\inst{1}\thanks{Present address: Inter University 
	Center for Astronomy \& Astrophysics, Post Bag 4 Ganeshkhind, 
	Pune 411 007, India: dmitra@iucaa.ernet.in} \and R. Ramachandran\inst{2}}
\offprints{dmitra@iucaa.ernet.in}
\institute{Raman Research Institute, Bangalore - 560 080, India. \email{dmitra@iucaa.ernet.in} \and 
	Stichting ASTRON, Postbus 2, 7990 AA Dwingeloo, The Netherlands. \email{ramach@astro.uva.nl}}
\authorrunning{D. Mitra \& R. Ramachandran}
\titlerunning{Scatter broadening of pulsars towards the Gum nebula}
\date{Received 24 November 2000 / Accepted 7 February 2001}
\abstract{We have measured the scatter broadening of pulsars in the direction of
the Gum nebula. For the first time, our observations show clear variations of
scattering properties across the Gum nebula. The IRAS-Vela shell is shown to be
a high scattering region. Our revised estimations of distances to these pulsars
are consistently less by a factor of 2--3, which has very important consequences
for the deduced values of radio luminosity and transverse velocity of pulsars.
}

\maketitle
\section{Introduction}\label{gum}
 The Gum nebula (first observed by Gum 1952, 1956) has the most extended
$\rm{H\alpha}$ emission observed in the sky. The nebula extends for about
$40^{\circ}$ in angular size centered around galactic longitude $l\sim
255^{\circ}$ and galactic latitude $b \sim -2^{\circ}$. The distance to the
nebula is found to be approximately $\sim 400 \rm{pc}$. Since the discovery of
the Gum nebula its origin has been an extremely controversial topic which still
has not been resolved.  There are several theories for the nature of the Gum
nebula and we briefly mention a few here. For a detailed review of the Gum
Nebula see Bruhweiler et al. (1983).

 The Gum Nebula appears extremely diffuse and faint in $\rm{H}\alpha$ thus
making it extremely difficult to estimate its size. One of the earliest
measurements gave its size to be as large as $75^{\circ} \times 45^{\circ}$
(Brandt et al. 1971).  Refined estimates of the size of the nebula using wide
field H$\alpha$ imaging is given by Sivan (1974), which restricts the size to
$\sim 36^{\circ}$.  Based on a spectroscopic study of ionized gas, Reynolds
(1976 a,b) proposed that the Nebula is a one million year old expanding gas
shell, originally produced by a supernova explosion, which is now being heated
and ionized by the massive stars $\zeta$ Puppis and $\gamma^{2}$
Velorum. According to Reynolds, the average density of the nebula is as large as
about 2 cm$^{-3}$. Assuming a diameter of about 250 pc, we can see that the
nebula can potentially introduce a dispersion measure of about 500 pc
cm$^{-3}$. 

According to Weaver et al. (1977), the stellar wind from $\zeta$ Puppis could be
strong enough to produce the observed Nebula, which is a shell. The shell is
formed by the interaction of the stellar winds from $\zeta$ Puppis and
$\gamma^{2}$ Velorum with the ambient interstellar medium. They also predict
soft X-rays from the hot interior, which is at a temperature of about
$10^{6}$K. Wallerstein et al. (1980), from a study of the interstellar gas
towards stars in the direction of the nebula, came to the conclusion that the
nebula is consistent with a model of the Gum Nebula as an HII region ionized by
OB stars and stirred up by multiple stellar winds. Chanot \& Sivan (1983), on
the basis of $60^{\circ}$-field $\rm{H\alpha}$ photographs, suggested that the
Gum Nebula is composed of two regions, one which is a circular main body with a
typical ring-like appearance of diameter of 36$^{\circ}$, and the other which
consists of faint diffuse and filamentary extensions which merge with the faint
$\rm{H\alpha}$ background. This idea supports the model of Reynolds (1976 a) for
an expanding $\rm{H\alpha}$ shell ionized by UV flux of $\zeta$ Puppis and
$\gamma^{2}$ Velorum. The origin of the shell structure is, however, uncertain.

From a detailed study of the Gum Nebula, Sahu (1992) came to the conclusion that
the nebula is a shell-like structure surrounding the Vela R2 association which
is at a distance of about 800 pc, while the shell-like structure near the Vela
OB2 association known as the IRAS Vela shell is at a distance of about 450
Kpc. This hypothesis crucially depended on the distance to $\zeta$ Puppis, as
this star is believed to be the primary source of ionization of the Gum Nebula,
which Sahu (1992) found to be $\sim$ 800 pc.  However the Hipparcos distance
estimates to $\zeta$ Puppis rules out the above scenario. Rajagopal (1999)
showed from the kinematics of the IRAS Vela shell that the Gum Nebula is either
inside or overlapping with the shell.

The various possible alternative scenarios as discussed above has left several
open ends in our understanding of the Gum Nebula. The electron density estimates
inside the Gum Nebula, as shown by Reynolds (1976a), and Wallerstein et
al. (1980), show large variations from 0.1 to 100 $\rm{cm^{-3}}$.  The
complicated structure of the nebula is also evident from the $\rm{H\alpha}$
images available (e.g.  Chanot \& Sivan, 1983).

Taylor \& Cordes (1993, hereafter TC93), as part of their galactic free electron
density distribution model, considered the Gum nebula as a separate component,
with an angular diameter of 30 degrees. They assumed that the number density of
free electrons is uniform all over this component (0.2 cm$^{-3}$). They also
assumed that the fluctuation parameter, which determines the amount of
scattering introduced by the medium, is zero. Though this is a drastic
assumption (as, for example, demonstrated by the scattering properties of the
Vela pulsar), as they state, this is mainly due to very poor constraints
available. In this work, we have done a systematic survey across the Gum Nebula
to measure the scatter broadening of pulsars due to the electron density
fluctuations, which should eventually help to model the scattering properties of
this region in more detail.

\section{Sample Selection and Observation}
We have observed 40 pulsars located in the galactic coordinate range
$250^{\circ}\; < \;l\;<\; 290^{\circ}$ and $-20^{\circ}\;<\;b\;<\;20^{\circ}$
with the Ooty Radio Telescope during March 1997. The Ooty Radio Telescope is a
semi-parabolic cylinderical array, whose dimensions are about 500 meters in
North-South, and 30 meters in East-West. It operates at a fixed centre frequency
of 327 MHz. It has 1056 dipoles arranged North-South along the focal line of the
semi parabolic cylinder, and as the result of which it is not sensitive to the
other (East-West) component of polarization (for further details 
refer Swarup et al 1971, Sarma et al 1975a., 1975b, Kapahi et al 1975).

\begin{figure}
\epsfig{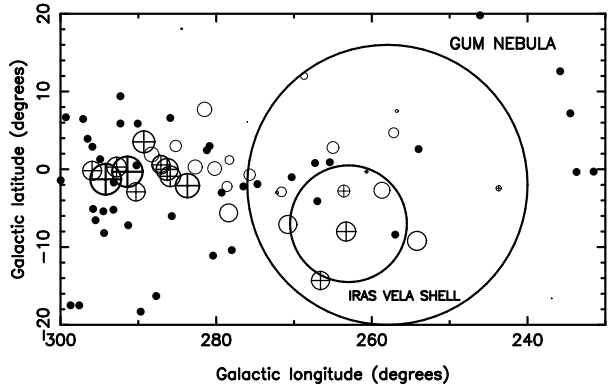}
\caption[]{Observed scatter broadening ($\tau_{\rm sc}$) of pulsars at 327 MHz
plotted as a function of the galactic coordinates. The circles with crosses
correspond to pulsars for which earlier $\tau_{\rm sc}$ measurements are
available. The open circles are the pulsars observed using the ORT. Size of
these symbols is proportional to $log(\tau_{\rm sc})$, where $\tau_{\rm sc}$ is
in milliseconds. Pulsars indicated as ``dots'' are those for which the scatter
broadening measurements are not available. The Gum nebula is indicated by the
larger circle with radius of $\sim 18^{\circ}$ with morphological center at
$l=258^{\circ}$ and $b=-2^{\circ}$. The smaller circle indicates the IRAS-Vela
shell, with a radius of $\sim 7.5^{\circ}$, with centre at $l=263^{\circ}$ and
$b=-7^{\circ}$ (adopted from Sahu 1992). Our sample consists also of pulsars
outside the ``main body'' of the Gum Nebula (Chanot \& Sivan 1983), as it is
possible that the faint diffuse and filamentary extensions are possibly part of
the nebula itself.}
\label{5gumregion}
\end{figure}

Although there are in total 48 known pulsars in this range, we selected only
those above a flux (at 400 MHz) of 5 mJy. The dispersion measures ($DM$) of
these pulsars are in the range 30 -- 306 pc cm$^{-3}$. Our sample also includes
pulsars for which scatter broadening measurements at other frequencies exist in
the literature.

To carry out these observations, we have used a pulsar receiver that was mainly
built for pulsar searches (Ramkumar et al. 1994). The pulsar receiver consists
of a 4-bit sampler (Analog-to-digital converter), which samples the incoming
signal voltage of bandwidth 8 MHz. The output of this is fed to an FFT
engine. The FFT produces 256-point complex spectra which are converted to power
spectra using look-up tables. The resultant power spectra are pre-integrated
over successive spans of $\sim 0.5$ msec, which is the final time resolution in
recorded data.  A block integration is done over a number of pre-integrated
samples for calculation of the running mean for each of the 256 frequency
channels. The running mean is finally subtracted from the pre-integrated data to
remove the effects of receiver gain variations.  The mean subtracted
pre-integrated data are then represented as a one-bit signal by recording the
sign bit and stored on magnetic tape. We observed each pulsar for 20 min, and 
compensated for the interstellar dispersion by offline analysis. The integrated
pulse profiles were obtained by folding the time series with the correct
rotation period of the pulsar.

\subsection{Measurement of scatter broadening}
Out of the 40 pulsars in our sample, only 21 were above our detection limit. For
each observed pulsar, we compensate for the interstellar dispersion by offline
software, and fold the time series with the exact expected rotation period, to
produce average pulse profiles.

The scatter broadening was then estimated by a least-square-fit to the average
pulse profile with the following model: The observed pulse profile function
$P(t)$ is the convolution of the intrinsic pulse profile (which is emitted by
the pulsar) $P_i(t)$ with (1) the impulse response characterising the scatter
broadening in the ISM $T(t)$, (2) the dispersion smearing function across the
spectral channel in the receiver $S(t)$, and (3) the instrumental response
function $I(t)$.
\begin{equation}
P(t) \;=\; P_i(t)\otimes T(t)\otimes S(t)\otimes I(t)
\label{equ1}
\end{equation}
where $\otimes$ denotes convolution. In a simple picture, where the scattering
material is assumed to be concentrated in a region whose thickness is very small
when compared to the distance between the observer and the pulsar, the impulse
response function is $T(t) = \exp(-t/\tau_{\rm sc})$. The whole procedure we have 
adopted to compute this $\tau_{\rm sc}$ is described in detail in 
Ramachandran et al. (1997).

The results of the fit are given in Table~\ref{5tablegum}. In
Table~\ref{5tablegum}, PSRs J0742--2822, J0745--5351 and J0835--4510 were not
observed by us; we have used the scattering measurements listed in the
literature (Roberts \& Ables 1982, Alukar et al. 1986). For those measurements
not done at 327 MHz, we have used the frequency scaling law $\tau_{\rm
sc}\propto \nu^{-4.4}$ (where $\nu$ is the observing frequency) to obtain the
scatter broadening value at 327 MHz. PSRs J0809--4753, J0837--4145 and J0840--5332
in Table~\ref{5tablegum} have had their scatter broadening reported earlier at
frequencies other than 327 MHz, but we have reobserved them. The values listed
for them in the table are from our measurements.


\begin{figure}
\epsfig{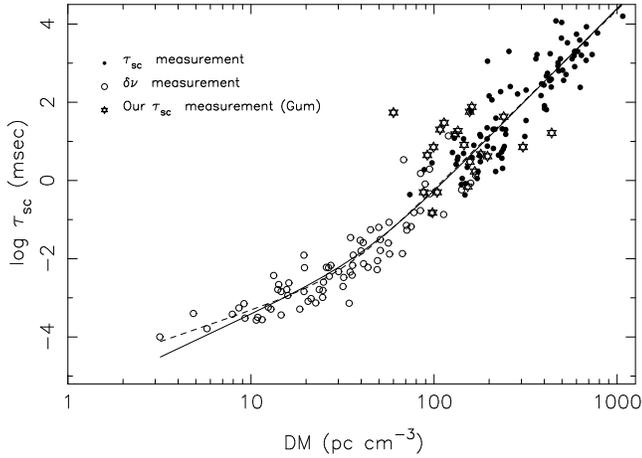}
\caption[]{A plot of $\tau_{\rm sc}$ vs $DM$ for 188 pulsars. The $\tau_{\rm sc}$
values have been scaled to 400 MHz. Stars in the plot represent our $\tau_{\rm
sc}$ measurements given in Table~\ref{5tablegum}. Open circles are from the
decorrelation band width measurements, using the relation $2\pi\nu\tau_{\rm
sc}=1$. Filled circles are the directly measured $\tau_{\rm sc}$ values. The
values of the open and filled circles are taken from the Princeton pulsar
catalogue (Taylor et al. 1993) and Ramachandran et al. (1997). The best fit 
models are plotted as lines. See text for details.}
\label{5gumscatplot}
\end{figure}

Figure~\ref{5gumregion} gives the spatial distribution of the pulsars across the
Gum nebula as a function of the galactic coordinates. Note that the distribution
is skewed to one side (mostly lying in the longitude range of 255$^\circ$ to
$275^{\circ}$) of the nebula marked by the large circle in the figure. This must
be primarily due to the fact that pulsars are mostly concentrated in the
galactic disk and at such galactic longitudes toward the Galactic anti-center
one does have significant contribution from the disk. Pulsars with high
$\tau_{\rm sc}$ appear to lie behind the IRAS vela shell, as marked by the small
circle in the figure.

\section{Discussion} 
\label{disgum}
In Figure \ref{5gumscatplot} we plotted the values of scatter broadening
$\tau_{\rm sc}$ for the whole pulsar population including our new results as a
function of dispersion measure ($DM$). A function of the form as given by
$\tau_{\rm sc}\;= A~DM^{\alpha}~(1+B~DM^{\gamma}) \lambda^{4.4}$ msec, as
discussed by Ramachandran et al (1997), is shown as a fit to the data points,
where $\lambda$ is the wavelength in meters.  On inclusion of our new
measurements the fit does not seem to change significantly. The dotted curve in
the figure corresponds to $A = 4.5 \times 10^{-5}$, $B = 3.1\times 10^{-5}$, 
$\alpha = 1.6$, $\gamma =
3$. The solid curve is modelled by fixing $\alpha = 2.2$ which is the expected
dependence from Kolmogorov spectrum, thus giving $A = 8.4 \times 10^{-6}$, $B =
8.3 \times 10^{-5}$ and $\gamma = 2.5$. The term $(1+B~DM^{\gamma})$ should
provide a useful description of the apparent mean dependence of the turbulence
level on $DM$. The scatter around the mean trend may be understood as due to
possible existance of isolated regions of enhanced scattering in the line of
sight, and consequent failure of the assumption that the scatterer is half way
down the line of sight.

Though PSR J0924--5814 seems to have a large deviation compared to the mean
Kolmogorov line, the error in the estimate of $\tau_{\rm sc}$ for this pulsar is
more than 100\% (refer table~\ref{5tablegum}) due to poor signal-to-noise ratio
of the integrated pulse profile.

\begin{figure}
\epsfig{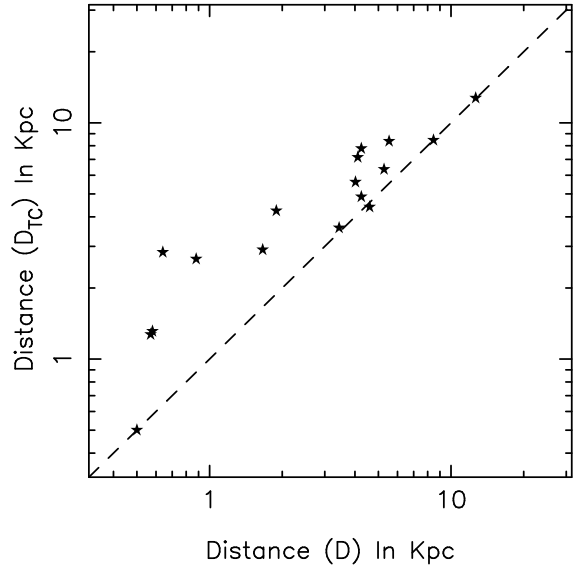}
\caption[]{ A comparison between the distance $D_{TC}$ obtained from 
the TC93  
model and the distance (D) calculated using our method is shown by the 
star symbol in the above figure. The dashed line correspond to the case $D = D_{TC}$.
Note that for most of the cases the model distances 
are overestimated (see text for more details).}
\label{fig:dgraph}
\end{figure}

\begin{table*}
\begin{center}
\caption[]{Summary of the measured scattering properties. Columns 1 -- 10
indicate (1) Pulsar name, (2) $DM$, (3) measured scatter broadening, $\tau_{\rm
sc}$, (4) intrinsic width, $w_i$ (5) error in $(\tau_{\rm sc}+w_i)$, (6) number
of degrees of freedom, (7) normalised $\chi^2$ values, (8) fluctuation
parameter, (9) derived distance, and (10) distance given by TC93
model. $^1$Those for which $\tau_{\rm sc}$ values are obtained from earlier
measurements: these were not observed by us. $^2$Those for which $\tau_{\rm sc}$
earlier measured values are available, but have also been reobserved by
us. $^{\dagger}$Those for which proper motion measurements are already
available.}
\begin{tabular}{l|c|c|c|c|c|c|c|c|c}\hline
{\bf Name} & {\bf $DM$} & {\bf $\tau_{\rm sc}$} & {\bf $w_{i}$ } & {\bf $\Delta
(\tau_{\rm sc} + w_{i})$} & {\bf $N_{dof}$} & {\bf $\chi^{2}$} & {\bf $F$} & {\bf
Distance} & {\bf D$_{TC}$} \\
 & {\bf (pc cm$^{-3}$)} & {\bf (msec)} & {\bf (msec)} & {\bf (msec)} & & & &
{\bf (kpc)} & {\bf (kpc)} \\
\hline
$^{\dagger}$J0738--4042       &160.8  & 76  $\pm$ 3   &             &     &&&  6.3 & 4.5 &   $>$11.03\\
$^{\dagger}$J0742--2822$^{1}$ &  73.7 & 1           &             &     & && 0.8 & 0.65 & 1.89  \\
J0745--5351$^{1}$ & 122.3 & 60           &             &     &&&  7.0    & 4.11& $>$ 7.14\\
J0809--4753$^{2}$ & 228.3 & 79  $\pm$ 18  &9   $\pm$10  & 8   &697  &1.03&5.9 & $>$12.65& $>$ 12.72\\
J0820--4114       & 113.4 & 30  $\pm$ 11  &42  $\pm$14  & 9   &216  &1.23&6.6 & 0.64& 2.83\\
$^{\dagger}$J0835--4510$^{1}$ & 68.2  & 8            &             &     &      &    &  6.3 & 0.50& 0.50\\
J0837--4135$^{2}$ & 147.6 &  1  $\pm$ 1   &9   $\pm$3   & 2   &597  &3   &0.0 & 1.89 & 4.24\\
J0840--5332$^{2}$ & 156.5 & 57  $\pm$ 11  &13  $\pm$10  & 10  &218  &1.1 &5.3 & 4.25& 7.78\\
J0846--3533       & 91.1  & 4   $\pm$ 5   &44  $\pm$6   & 5   &135  &2.1 &1.7 & 0.58& 1.31\\
J0855--3331       & 87.7  & 1   $\pm$ 3   &24  $\pm$1.5 & 1   &595  &1.3 &0.0 & 0.57& 1.27\\
J0904--4246       & 189   & 8   $\pm$ 4   &21  $\pm$4   & 3   &397  &0.99&0.8 & 4.6 & 4.40\\
J0905--5127       & 189   & 4   $\pm$ 7   &11  $\pm$8   & 5   &218  &0.88&0.5 & 5.54& 8.35\\
J0907--5157       & 104   & 1   $\pm$ 13  &32  $\pm$10  & 10  &228  &0.9 &0.15& 0.88& 2.65\\
J0924--5302       & 152.9 & 1   $\pm$ 1   &6   $\pm$0.8 &  1  &195  &1.1 &0.15& 4.02& 5.61\\
J0924--5814       & 60.0  & 55  $\pm$ 65  &1   $\pm$50  &45   &238  &1.0 &    &     &2.01\\
J0934--5249       & 99.4  & 7   $\pm$ 13  &21  $\pm$10  &5    &197  &1.1 &1.3 & 1.66&2.90\\
J0942--5552       & 180.2 & 5   $\pm$ 1.5 &7   $\pm$2   &2    &237  &1.1 &0.3 & 5.28& 6.35\\
J0952--3839       & 167   & 2   $\pm$ 12  &48  $\pm$10  &8    &227  &0.91&0.25& $>$8.46&$>$ 8.44\\
J0955--5304       &156.9 & 3    $\pm$ 1   &5   $\pm$1   &1    &122  &1.2 &0.67& 4.25& 4.86\\
J1001--5507       & 130.6 & 15  $ \pm$ 2  &13  $\pm$2   &2    &197  &1.1 &10.7& 3.44& 3.59\\
J1003--4747       &98.1   & 0   $\pm$ 4   &9   $\pm$4   &3    &197  &1.2 &    &     &3.44\\
J1017--5621       &439.1  & 16  $\pm$ 36  &8   $\pm$25  &25   &228  &0.97&    &     &11.77\\
J1042--5521       &306    & 7   $\pm$ 10  &29  $\pm$10  &10   &238  &1.1 &    &     &6.95 \\
J1046--5813       &240.2  & 43  $\pm$ 50  &2   $\pm$50  &50   &238  &1.03&    &     &4.8 \\
J1059--5742       &107.9  & 2   $\pm$ 3   &20   $\pm$7   &5   &237  &1.2 &     &    &2.74\\
\hline
\end{tabular}
\label{5tablegum}
\end{center}
\end{table*}

The presence of the Gum nebula has been invoked explicitly in models estimating
pulsar distances and free electron density distribution in the Galaxy (e.g.
Bhattacharya et al 1992, TC93 and references therein). TC93 modelled the Gum
nebula as a sphere of 130 pc radius at a distance of 500 pc, with a uniform
electron density of 0.25 cm$^{-3}$, and the density falls off as an one-sided
Gaussian with an r.m.s. of 50 pc. They also assumed that the fluctuation
parameter, which is defined as
\begin{equation}
F\;=\; \frac{\zeta\epsilon^2}{\eta}\;\left(\frac{l}{{\rm 1\;pc}}\right)^{-2/3}\;,
\end{equation}
\noindent
is zero. Here, $\zeta$ is the normalised variance of large-scale electron
density fluctuations, $\epsilon^2$ is the corresponding quantity for small-scale
density variations, $\eta$ the volume filling factor for ionised regions, and
$l$ the outer scale limit for an assumed Kolmogorov power law of the density
turbulence of free electrons. The assumption that $F$ is zero means that the
nebula contributes to dispersion measure, but not to the scattering
properties. As the authors state, this is mainly due to the very poor knowledge
of the scattering properties of pulsars in that direction.

From our observation, it is obvious that many parts of the nebula significantly
contribute to enhanced scattering of pulsar signals. This issue was addressed by
Deshpande \& Ramachandran (1998) in detail, where they explicitly showed that in
order to explain the enhanced scattering observed for PSR J0738--4042, which is
a pulsar in the Gum Nebula (refer table~\ref{5tablegum}), one has to increase
$n_{e}$, and adopt values of fluctuation parameter almost equal to that of the
spiral arm. We have used a similar method as suggested by Deshpande \&
Ramachandran (1998), and have obtained the distances and the fluctuation
parameters for the pulsars behind the Gum Nebula under the framework of the TC93
model. which we briefly describe below.

Deshpande \& Ramachandran (1998) showed that if the distance to a dominant
discrete scatterer is known, then it is possible to use only the $DM$ and the
$\tau_{\rm sc}$ measurements to constrain $F$ and distances to pulsars lying
behind the scattering region. With the available data on the Vela pulsar and
J0738--4042, they find similar values of $n_{e}$ and $F$, of about 0.32
cm$^{-3}$ and 6.3, respectively. This helped constrain the distance of
J0738--4042 to 4.5 kpc (as opposed to $>$11 kpc by TC93). As they suggest, it
seems reasonable to characterise a major part of the Gum nebula with $n_{e} =
0.32~ \rm{cm^{-3}}$ and a fluctuation parameter $F$ = 6.3. In a simple exercise,
for various lines-of-sight in the Gum Nebula we have applied the above technique
to obtain $F$ and distances to pulsars, keeping $n_{e}$ fixed at 0.32. The
values of $F$ and distances (in kpc) obtained is given in
table~\ref{5tablegum}. Though these values are not unique (as different
combinations of $n_{e}$ and $F$ can match the observed values of $DM$ and
$\tau_{\rm sc}$), as Deshpande \& Ramachandran (1997) show, it is not
unreasonable. It is interesting to note that most of the pulsars lying in the
IRAS Vela shell seems to be consistent with a fluctuation parameter of $\sim
6.5$, while the other regions in the Gum Nebula have an $F$ of only
$\sim$0.5. This, at least intuitively, indicates that the IRAS-Vela shell is a
different entity with different fluctuation properties. Note that we have not
attempted any modelling of the region outside the Gum Nebula. The value of $F$
for PSR J1001--5507 is 10.7, which we believe is too high to be associated with
the nebula. For PSR J0924--5814, we get a value of 25 which is unreasonable, and
this is due to the poor estimation of $\tau_{\rm sc}$. Thus, we reject this
pulsar in our analysis.

The distances we obtain from our procedure seem to be consistently different
from the values obtained by the TC93 model (as shown in Fig. \ref{fig:dgraph}).
It is worth noting that out of this list in the table, proper motion
measurements are already available for those pulsars marked with a dagger
($\dagger$). For them, on the average, our analysis makes a difference of 2--3
in the estimated value of transverse velocity.

As an extension of the present study to a more detailed one, it would be
interesting to establish the electron density variation of the Gum Nebula along
the various lines-of-sight of these pulsars. As found by Reynolds (1976b), the
number density in the Gum nebula can in principle vary by many times, with an
average density of about 2 cm$^{-3}$. This can potentially introduce dispersion
measures of the order of 500 pc cm$^{-3}$. Estimates of density variations can
be obtained from detailed H$\alpha$ studies, which involves measuring emission
measures in the line-of-sight to pulsars. The ratio of the emission measure and
the $DM$ can be used to estimate the electron densities. Although such estimates
are available (Reynolds, 1976a), they are not sufficient for the entire set of
lines-of-sight observed. Further, as we have estimated the scatter broadening of
only a subset of pulsars in this region, there remains a significant fraction of
pulsars for which such measurements are not available (as clearly seen in figure
\ref{5gumregion}). With a more sensitive instrument, it should be possible to
enlarge the sample of scatter broadening measurements, giving further clues
about the electron density distribution.

\section*{Acknowledgement} We would like to thank Dr. A. A. Deshpande of the Raman 
Research Institute for his invaluable help and stimulating discussions. We would 
also like to thank V. Balasubramanium for providing us with telescope time and help 
during the observations in Ooty. We would like to thank Bilal and Mangesh for their 
kind help during the observations.

\beb
\bi Alurkar, S.K., Slee, O.B. \& Bobra, A.D. 1986, AJP, 39, 433.
\bi Bhattacharya D., Wijers R. A. M. J., Hartman J. W. \& Verbunt F. 1992, A\&A, 254, 198. 
\bi Brandt, J. C., Stecher, T. P., Crawford, D. L. \& Maran, S. P., 1971, ApJ. Letters, 163, L99.
\bi Bruhweiler F. C., Kafatos M. \&  Brandt J. C., 1983, Comments on Modern
    Physics, Part C - Comments on Astrophysics (ISSN 0146-2970), 10, 1.
\bi Chanot, A. \& Sivan, J. P., 1983, A\&A, 121, 19.
\bi Deshpande A. A. \& Ramachandran R, 1998, MNRAS, 300, 577.
\bi Gum, C. S., 1952, Observatory, 72, 151.
\bi Gum, C. S., 1956, Observatory, 76, 150.
\bi Gwinn, C. R., Bartel, N. \& Cordes, J. M., 1993, ApJ, 410, 673.
\bi Kapahi, V. M., Damle, S. H., Balasubramanian, V. \& Swarup, G., 1975, J. IETE, 21, 117.
\bi Ramkumar P. S., Prabu T., Girimaji Madhu, \& Markandeyulu G, 1994, JApA, 15, 343.
\bi Rajagopal, J., 1999, Ph.D. thesis, Raman Research Institute, Bangalore.
\bi Reynolds, R. J., 1976a, ApJ, 203, 151.
\bi Reynolds, R. J., 1976b, ApJ, 206, 679.
\bi Roberts, J. A. \& Ables, J. A., 1982, MNRAS, 201, 1119.
\bi Ramachandran R., Mitra D., Deshpande A. A., McConnell D. M. \& Ables J. G., 1997, MNRAS, 290, 260.
\bi Sivan, J. P., 1974, Astr. Astrophys. Suppl., 16, 163.
\bi Sahu, M. S., 1992, Ph.D. Thesis, University of Groningen.
\bi Sarma, N. V. G., Joshi, M. N.\& Ananthakrishnan S., 1975b, J. IETE, 21, 107.
\bi Swarup, G., Sarma, N. V. G., Joshi, M. N., Kapahi, V. K., Bagri, D. S., Damle, S. H.,
Ananthakrishnan, S., Balasubramanian, V., Bhave, S. S. \& Sinha, R. P., 1971,
Nature Phys. Science, 230, 185.
\bi Sarma, N. V. G., Joshi, M. N., Bagri, D. S., \& Ananthakrishnan, S., 1975a, J. IETE, 21, 110.
\bi Taylor, J.H., \& Cordes, J.M. 1993, ApJ,  411, 674.
\bi Taylor, J. H., Manchester, R. N. \& Lyne, A. G., 1993, ApJS, 88, 529.
\bi Wallerstein, G., Silk, J. \& Jenkins, E. B., 1980, ApJ, 240, 834.
\bi Weaver, R., McCray, C. T., C. T., Castor, J., Shapiro, P. \& Moore, R., 1977, ApJ, 218, 377.
\eeb

\end{document}